\documentclass[aip,reprint]{revtex4-2}

\usepackage{graphicx}
\usepackage{dcolumn}
\usepackage{bm}

\usepackage[utf8]{inputenc}
\usepackage[T1]{fontenc}
\usepackage{mathptmx}
\usepackage{etoolbox}
\usepackage{natmove}
\usepackage[caption=false]{subfig}
\usepackage{chemformula} 
\usepackage{physics}
\usepackage[capitalise]{cleveref}
\usepackage[cal=cm]{mathalfa}
\usepackage[acronym]{glossaries}
\glsdisablehyper

\usepackage[caption=false]{subfig}
\newacronym{md}{MD}{Molecular Dynamics}
\newacronym{qm}{QM}{Quantum Mechanics}
\newacronym{csdf}{CSDF}{Center of the Spin Density Function}
\newacronym{vebe}{VEBE}{Vertical Electron Binding Energy}
\newacronym{dft}{DFT}{Density Functional Theory}
\newacronym{oml}{OML}{Orbital Motion-Limited}

\newcommand{\cvp}{\varepsilon_{\mathrm{0}}}

\newcommand{\E}{\mathcal{E}}

\renewcommand{\Phi}{\varPhi}

\begin{document}

\title{Equilibrium Elongation of Charged Nanoclusters}

\author{Efstratios M. Kritikos}
\email{emk@caltech.edu}
\affiliation{Department of Applied Physics and Materials Science, California Institute of Technology, Pasadena, CA 91125, United States}
\affiliation{Department of Physics, University of Oxford, Parks Road, Oxford, OX1 3PU, United Kingdom}

\author{William A. Goddard III}
\affiliation{Materials and Process Simulation Center, California Institute of Technology, Pasadena, CA 91125, United States}
\affiliation{Department of Applied Physics and Materials Science, California Institute of Technology, Pasadena, CA 91125, United States}

\author{Paul M. Bellan}
\affiliation{Department of Applied Physics and Materials Science, California Institute of Technology, Pasadena, CA 91125, United States}

\begin{abstract}
When internal electrostatic stresses from electron repulsions overcome surface tension, charged nanoclusters elongate. However, the exact shape they reach at equilibrium has been unknown. We present an analytical model for the equilibrium structure of charged ellipsoidal nanoclusters via a free energy analysis. Energy maps are constructed that determine the equilibrium states as a function of cluster size, net charge, surface tension, and local curvature. The model is compared with molecular dynamics simulations of ice nanoclusters having solvated electrons. Despite the discreteness of the simulated atoms and electrons, both the predictions of the continuum analytical model for elongation and the equilibrium elongated shape show good agreement. The model could therefore predict nanocluster shape evolution in astrophysical and laboratory complex plasmas and atmospheric aerosols.
\end{abstract}

\maketitle

Charged nanoclusters, primarily composed of ice, naturally occur in astrophysical and laboratory plasmas \cite{Shukla2015}. Experimental measurements have found that interstellar matter is elongated, causing starlight to become polarized \cite{Davis1951}. These elongated particles align with the magnetic field lines in astrophysical environments, such as in star-forming regions and in protoplanetary disks, allowing the direct measurement of the magnetic field strength \cite{Tram2022}. Understanding the morphology of these particles is essential in predicting their magnetic properties and thus their Larmor precession and magnetic relaxation. 

The origin of this elongation in astrophysical plasmas has yet to be identified. We recently proposed that elongation begins at nanoscale sizes due to internal electrostatic stresses originating from the negative charging of the nanocluster in a plasma \cite{Kritikos2025}. After an elongated nanocluster is created, it can continue to grow into elongated shapes \cite{Nicolov2024} due to the accumulation of polar water molecules at the tips of the ellipsoidal clusters, driven by the strong electric field gradient there \cite{Bellan2020}. Therefore, the growth of these particles is directly linked to their initial morphology.

Charged nanoclusters are also ubiquitous in Earth's atmosphere. In noctilucent clouds, experimental evidence suggests that ice particles are nanosized and elongated \cite{Hervig2009}. Additionally, precipitation particles with a highly negative charge exist in thunderstorms \cite{Gaskell1977}. Charged nanoparticles also have biomedical applications, such as in cellular uptake and diagnostics \cite{Fang2009}. In these biological fluids, the suspended nanoparticles can aggregate and, due to electrostatic repulsions, lead to elongated large clusters \cite{Zhang2010}. Charged nanoclusters also form spontaneously in weakly ionized plasmas used for semiconductor fabrication, and unless removed, their deposition can degrade the chips \cite{Selwyn1995}. Understanding the properties of these particles will help develop effective technologies to remove them.

The deformation of charged droplets due to electrostatic stresses was first formulated by Rayleigh \cite{Rayleigh1882}. When electrostatic stresses overcome surface tension stresses, the liquid droplet deforms into an ellipsoid (prolate spheroid) \cite{Grimm2005,Giglio2008}, and if the stresses become too strong, then Taylor cones are formed, i.e., jets of charged particles \cite{Taylor1964}. The stability criterion of charged droplets has been extensively investigated using theoretical and numerical approaches\cite{Muratov2016,Consta2022}. In these studies, nanoscale corrections to the surface tension are typically neglected, as the focus is on macroscale droplets. Moreover, focus is given to the deformation of liquid clusters, rather than low-temperature solid ones, which are more likely to form protrusions instead of reaching elongated shapes\cite{Kritikos2025}. Recently, we developed an analytical model for the deformation of charged ellipsoidal clusters based on a local stress analysis at the tip while also considering nanoscale surface tension corrections \cite{Kritikos2025}. Even though the analysis explained when a charged ellipsoidal cluster elongates, the exact equilibrium shape of the cluster remains unknown.

In this work, we present an analytical model for the equilibrium shape of charged nanoclusters as they deform under constant charge or constant electric potential. Also, we perform \gls{md} simulations of charged ice nanoclusters to validate the analytical model. We focus on nanoclusters composed of ice, as they are directly relevant to astrophysical plasmas and atmospheric aerosols, but the analytical model is expected to apply to any type of cluster.

\begin{figure}
    \centering
    \includegraphics[width=0.6\linewidth]{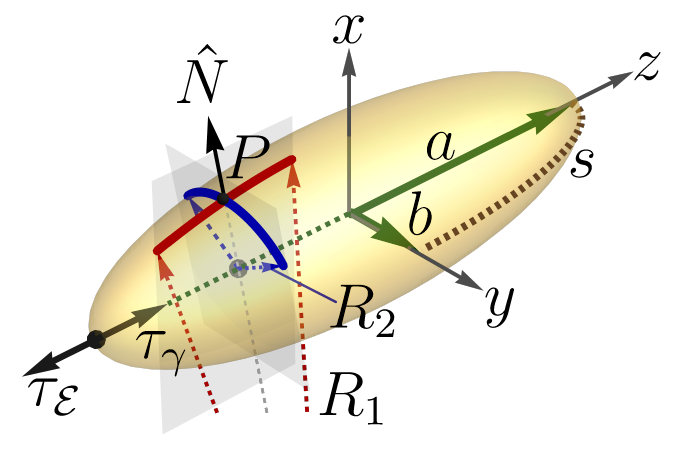}
    \caption{Characteristic dimensions of the ellipsoidal nanocluster.}
    \label{fig:ellips_params}
\end{figure}

The equilibrium aspect ratio of charged nanoclusters is determined here via their free energy. For our analysis, the elongated nanoclusters are approximated as ellipsoids (prolate spheroids) with $a$ and $b$ being the major and minor radii. We also define the aspect ratio, $\alpha=a/b$, and the radius $R$ of a same-volume sphere, which follows from $R^3=a b^2$. Another property of the ellipsoid is the total curvature of the surface $K=R_1^{-1}+R_2^{-1}$, where $R_1$ and $R_2$ are the two principal radii at a point on the surface. The parameters are shown in \cref{fig:ellips_params}. From the above definitions, the relations $a=\alpha^{2/3}R$ and $b=\alpha^{-1/3}R$ follow.

The total Gibbs free energy of a charged ellipsoidal nanocluster depends on its surface tension free energy $G_\gamma$ and the electrostatic free energy $G_\E$, i.e., 
\begin{equation}
    G= G_\gamma + G_\E. 
\end{equation}
We use the Gibbs free energy of a same-volume sphere as the reference point, and so define
\begin{equation}
    \Delta G=G_{ell}-G_{sph}.
\end{equation}
Therefore, the total change in free energy from a sphere becoming a same-volume ellipsoid is given by
\begin{equation}
\Delta G=\Delta G_\gamma + \Delta G_\E.
\label{eq:delta_G}
\end{equation}
A cluster with $\Delta G>0$ is less stable compared to a sphere, and thus at least one state with lower energy (i.e., that of a sphere) exists. A local minimum on the $\Delta G$ curve, $\Delta G_{l,\min}$, indicates a local equilibrium shape for the cluster, whereas a local maximum denotes an unstable configuration. The global minimum, $\Delta G_{g,\min}$, of the map indicates the most stable structure for the given conditions. When $\Delta G=0$, the ellipsoidal and spherical nanoclusters have the same energy, indicating a sign change along the $\Delta G$ curve for a given $R$ and thus that multiple local minima exist.

First, we derive the free energy due to surface tension. As the cluster size decreases, surface tension deviates from the planar limit, and nanoscale corrections must be included. Tolman introduced a radius-dependent correction to the surface tension for small spherical droplets \cite{Tolman1949}. Later, Helfrich presented a surface tension formulation for an arbitrary-shaped surface based on the total curvature \cite{Helfrich1973}. Combining the two formulations, Tolman's correction can be written as
\begin{equation}
    \gamma = \dfrac{\gamma_\infty}{1+K\delta},
    \label{eq:gamma}
\end{equation}
where $\delta$ is the Tolman length, a material-dependent parameter. For water, there is no universally accepted $\delta$ value. The values vary significantly among studies, ranging\cite{Kritikos2025} from $-2.0$ to 1.5 {\AA}, effectively predicting different $\gamma$ trends. The Tolman length is also affected by the nanocluster temperature and surface charge. For the theoretical results below, we use $\delta=-0.8$ \AA, which was computed\cite{Kritikos2025} via \gls{qm} simulations for water clusters with solvated electrons at 100 K.

\begin{figure*}[t]
    \centering

    \includegraphics[width=0.48\linewidth]{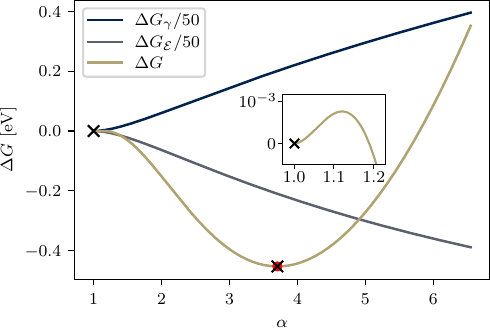}\hspace{6pt}%
    \includegraphics[width=0.48\linewidth]{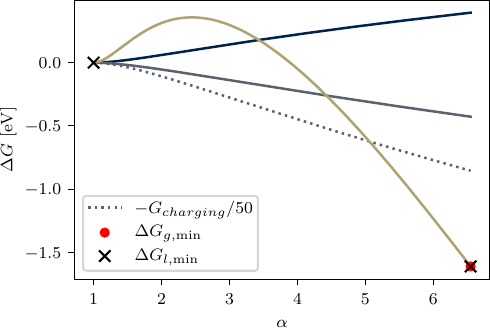}%
\vspace{-10pt}
\subfloat[Elongation at constant $Q$]{%
    \includegraphics[width=0.5\linewidth]{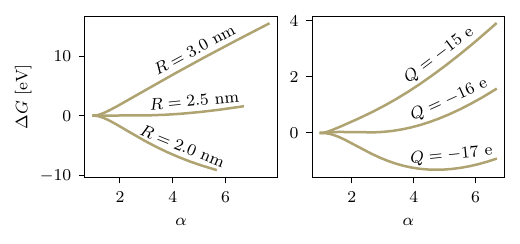}}%
\subfloat[Elongation at constant $\Phi$]{%
    \includegraphics[width=0.5\linewidth]{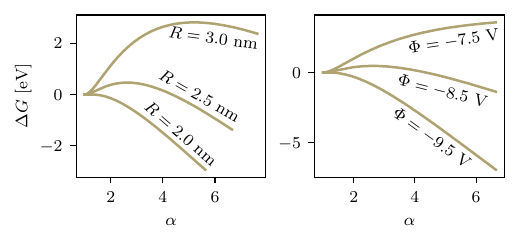}}

    \caption{Individual components of $\Delta G$ at (a) $Q=-16$ e and (b) $\Phi=-8.5$ V deformation for $\gamma_{\infty}=76$ mN/m, $\delta=-0.8$ {\AA}, $R=2.45$ nm. The $\Delta G_{l,\min}$ and $\Delta G_{g,\min}$ are also indicated. For the bottom row graphs, the aforementioned conditions are assumed unless otherwise specified by the curve labels.}
    \label{fig:energy_indiv}
\end{figure*}

The surface tension energy of a surface having area $A$ is 
\begin{equation}
    G_\gamma=\iint_A\gamma \dd A.
    \label{eq:G_gamma}
\end{equation}
The ellipsoid surface in \cref{fig:ellips_params} results from the revolution of an ellipse around the $z$ axis. Then, $A=\iint r \dd s_r$, where $s_r$ is the path of the ellipse, and thus $A = \int_0^{2\pi}\int_{-a}^a r \sqrt{1+( \dd r/\dd z )^2} \dd z \dd \phi$. The ellipsoid is defined by $r^2/b^2+z^2/a^2=1$. Therefore, $\dd r /\dd z=- b z/(a \sqrt{a^2-z^2})$ and so $A$ can be written as 
\begin{equation}
    A = \alpha^{-2/3} R^{-1} [ \alpha^{4/3}R^2 + (\alpha^{-2}-1) z^2]^{1/2}.
    \label{eq:area}
\end{equation}
Combining \cref{eq:gamma,eq:G_gamma,eq:area}, the surface tension free energy of an ellipsoid becomes $G_{\gamma,ell}=2\pi b \int_{-\alpha^{2/3}R}^{\alpha^{2/3}R} \gamma \alpha^{-2/3}R^{-1} [\alpha^{4/3}R^2+(\alpha^{-2}-1)z^2]^{1/2}\dd z$. 

For the surface tension given in \cref{eq:gamma}, we need to calculate the principal radii $R_1$ and $R_2$. The derivation, given in Section~S1.1, shows that
\begin{equation}
R_1=\alpha^{-1/3}R^{-2}[\alpha^{4/3}R^2+(\alpha^{-2}-1)z^2]^{3/2}
\end{equation}
and 
\begin{equation}
R_2=\alpha^{-1}[\alpha^{4/3}R^2+(\alpha^{-2}-1)z^2]^{1/2}.    
\end{equation}
The total curvature is thus given by
\begin{equation}
    K=\dfrac{\alpha^{1/3}R^2+\alpha[\alpha^{4/3}R^2+(\alpha^{-2}-1)z^2]}{[\alpha^{4/3}R^2+(\alpha^{-2}-1)z^2]^{3/2}}.
    \label{eq:tot_curvature}
\end{equation}
Using \cref{eq:gamma,eq:G_gamma,eq:area,eq:tot_curvature}, the surface tension energy of an ellipsoid becomes
\begin{equation}
    G_{\gamma,ell}=\dfrac{2\pi \gamma_\infty}{\alpha}\int_{-\alpha^{2/3}R}^{\alpha^{2/3}R}\dfrac{f^2}{f^{3/2}+\delta\alpha^{1/3}R^2+\delta \alpha f}\dd z,
\end{equation}
where $f(z)=\alpha^{4/3}R^2+(\alpha^{-2}-1)z^2$. The $z$ integral is evaluated numerically. For a sphere (i.e., $\alpha \rightarrow 1$), the surface tension energy reduces to $G_{\gamma,sph}=4\pi R^3 \gamma_\infty/(R+2\delta)$.

We next derive the free energy due to electrostatic repulsions. We assume that the nanocluster is a conductor. As shown later, MD simulations of nonconducting charged water clusters agree well with this assumption. The free energy of a charge distribution in a conductor is 
\begin{equation}
G_\E = \dfrac{1}{2}\iint_A \sigma \Phi dA, 
\end{equation}
where $\Phi=Q/C$. The capacitance of an ellipsoid is \cite{Shumpert1972}
\begin{equation}
    C_{ell}=\dfrac{8\pi \cvp \alpha^{2/3}R\sqrt{1-\alpha^{-2}}}{\ln(\frac{1+\sqrt{1-\alpha^{-2}}}{1-\sqrt{1-\alpha^{-2}}})}
\end{equation}
and of a sphere is $C_{sph}=4\pi \cvp R$. Combining the above terms, the electrostatic energy becomes
\begin{equation}
G_{\E,ell}=\dfrac{Q^2 \ln{\frac{1+\sqrt{1-\alpha^{-2}}}{1-\sqrt{1-\alpha^{-2}}}}}{16\pi \cvp \alpha^{2/3} R \sqrt{1-\alpha^{-2}}}.
\label{eq:G_Eell}
\end{equation}
For a sphere, \cref{eq:G_Eell} simplifies to $G_{\E,sph}=Q^2/(8\pi \cvp R)$.

The free energy $\Delta G$ can differ, depending on how fast the cluster deforms with respect to how fast it gets charged. If the elongation timescale, $t_d$, which for low-temperature nanoclusters is governed by surface diffusion\cite{Kritikos2026}, is much smaller than the charging timescale $t_c$, i.e., $t_d\ll t_c$, then the cluster retains its initial charge as it elongates. If the nanocluster deforms slowly, such that $t_d \gg t_c$, then its charge will change according to its capacitance in order to maintain a surface equipotential. Elongation in a constant potential can occur for nanoclusters placed in electrosprays or plasmas, where they acquire a negative charge. In a plasma, according to the \gls{oml} theory \cite{Shukla2015}, the particle will become charged to the floating potential, which is typically $\sim2\times$ the electron temperature. Ellipsoids have a higher capacitance than spheres, and thus under constant potential their charge will increase with increasing $\alpha$. 

In that case, we have to consider the work required to increase the charge on the cluster as it elongates. This work represents a decrease in the energy of the system comprised of the cluster and the surrounding plasma, where the latter can be considered equivalent to a battery that provides the additional charge on the cluster. The work to bring charges of magnitude $\dd Q$ from the plasma to the cluster is $G_{charging}=\int_{Q_{sph}}^{Q_{ell}} Q\dd Q/C$. The capacitance is not constant, but the potential is, so using $Q=C\Phi$ and $\dd Q=\Phi \dd C$, the energy removed from the plasma as a result of the plasma charging the cluster is
\begin{equation}
    G_{charging} =\Phi^2 (C_{ell}-C_{sph}).
\end{equation}
The increase in cluster capacitance at constant voltage increases the energy of the cluster but decreases the energy of the plasma by a greater amount. Therefore, the change in free energy of the cluster-plasma system relative to that of a same-volume sphere having the original charge is   $\Delta G_{\mathcal{E}}=G_{\mathcal{E},ell}-G_{charging}-G_{\mathcal{E},sph}$ (an equivalent-circuit explanation is provided in Section~S1.2). A minus sign is in front of $G_{charging}$ because this work is a reduction in the free energy of the cluster-plasma system, as shown in the discussion leading to Eq.~(S25).  

\Cref{fig:energy_indiv} plots $\Delta G$, along with the local and global minima (i.e., $\Delta G_{l,\min}$ and $\Delta G_{g,\min}$, respectively), under constant $Q$ or constant $\varPhi$ conditions. At constant $Q$ (cf. \cref{fig:energy_indiv}a), surface tension increases $\Delta G_\gamma$ with increasing elongation, forcing an initially ellipsoidal cluster to become more spherical. Conversely, electrostatic energy $\Delta G_{\mathcal{E}}$ decreases as $\alpha$ increases. If the cluster is at an equipotential (cf. \cref{fig:energy_indiv}b), electrostatic energy decreases even further by $G_{charging}$. The competition between these energies determines the cluster's equilibrium shape. Results are plotted up to $\alpha$ values that give a minimum of a 2 {\AA} radius of curvature $R_c$ at the tip of the ellipsoid. This corresponds approximately to the size of a single water molecule, which is the smallest possible size for the water clusters studied here \cite{Kritikos2025}. 

For the particular case of \cref{fig:energy_indiv}, there is one global and two local minima both for elongation at constant $Q$ and constant $\Phi$. In the constant $Q$ elongation, the two local minima are separated by a small energy barrier as shown in the inset in \cref{fig:energy_indiv}a. This suggests that a cluster can be trapped in either equilibrium shape (spherical or elongated). The $\Delta G$ line also indicates that an initially spherical nanocluster must overcome an energy barrier at low $\alpha$ before reaching its maximum elongation. The bottom graphs in \Cref{fig:energy_indiv} show how varying $R$, $Q$, and $\varPhi$ affects $\alpha$. Larger clusters with smaller surface charge density experience weaker electrostatic stresses, moving their $\Delta G_{l,min}$ towards lower $\alpha$. A more detailed parametric analysis of $\Delta G$ is provided in Figs.~S2 and S3.

\begin{figure}[!tb]
    \centering
\subfloat[Elongation at constant $Q$\label{subfig:energy_maps_Q}]{%
    \includegraphics[width=\linewidth]{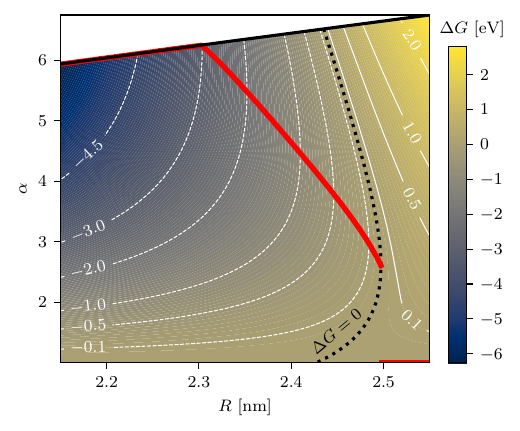}}
    
\subfloat[Elongation at constant $\Phi$\label{subfig:energy_maps_Phi}]{%
    \includegraphics[width=\linewidth]{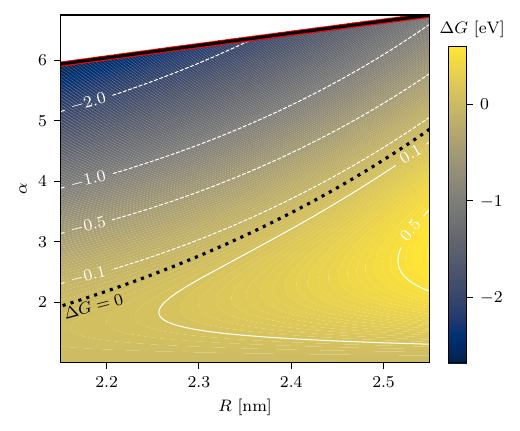}}
    \caption{$\Delta G$ maps at (a) $Q=-16$ e and (b) $\Phi=-8.5$ V deformation for $\gamma_{\infty}=76$ mN/m and $\delta=-0.8$ \AA. The red line corresponds to $\Delta G_{g,\min}$ and the dotted line to $\Delta G=0$. The white region denotes $R_c<2$ {\AA} values.}
    \label{fig:energy_maps}
\end{figure}

Using the above analysis, we can create $\alpha-R$ maps to find the equilibrium cluster shapes and thus possible evolution paths at different conditions; a case is plotted in \cref{fig:energy_maps}. The red line here shows the $R$ at which $\Delta G$ is at a minimum in \cref{fig:energy_indiv}; e.g., a value of $R=2.45$ nm on the red line of \cref{fig:energy_indiv}a corresponds to $\alpha=3.7$, which matches the red bullet point in \cref{fig:energy_indiv}a. For the constant charge deformation, shown in \cref{subfig:energy_maps_Q}, we can distinguish 3 main regions depending on the size of the cluster. At small sizes, the electrostatic stresses are too strong, pushing the cluster to elongate to its $R_c=2$ {\AA} physical limit. This large energy difference suggests rapid and violent deformation that could potentially lead to the breaking of the cluster. At large sizes, electrostatic stresses become insignificant, and the cluster cannot elongate. There is also an intermediate region at $R\in [2.3,2.5]$ nm where the equilibrium shape exists in valleys of $\alpha$ values. Moreover, cluster sizes of $R\in[2.4,2.5]$ nm have both a local minimum at $\alpha\rightarrow 1$ and a global minimum. These clusters will only reach their global minimum if they start elongating at smaller $R$ sizes (i.e., to the left of the dotted $\Delta G=0$ line in \cref{fig:energy_maps}a) and grow, or if a sufficient perturbation, e.g., strong thermal fluctuations, gives enough energy to overcome the barrier. 

For a constant potential deformation, the global minimum in \cref{subfig:energy_maps_Phi} corresponds to the smallest allowed tip curvature. The reason for this is that the capacitance of an ellipsoid increases with increasing $\alpha$, thus allowing for more charge to be collected by the cluster and thus stronger electrostatic stresses to develop. Note also that the positive $\Delta G$ barrier increases both in magnitude and $\alpha$ spread as $R$ increases.

The analysis of \cref{eq:delta_G}, which considers all points along the surface, predicts a different equilibrium $\alpha$ compared to when stresses at a single point, e.g., the tip, of the ellipsoid are evaluated \cite{Kritikos2025}. The reason is that even though stresses dominate at the ellipsoid tip, it is the collective contribution of all surface points that determines the equilibrium shape. For example, a 2.4 nm cluster in the \cref{fig:energy_maps}a conditions will stop elongating at $\alpha=1.4$ according to a stress evaluation at the ellipsoid tip but will reach $\alpha=4.6$ according to its minimum free energy. This indicates that to predict the equilibrium shape it is necessary to consider all stress contributions on the surface.

\begin{figure*}[!ht]
    \centering
    
    \begin{minipage}[c]{0.48\linewidth}
        \centering
        \subfloat[]{%
            \includegraphics[width=\linewidth]{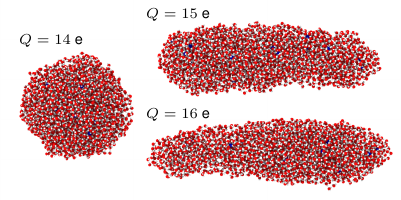}
            \label{subfig:md_screenshots}%
        }\\ 
        \subfloat[]{%
            \includegraphics[width=0.6\linewidth]{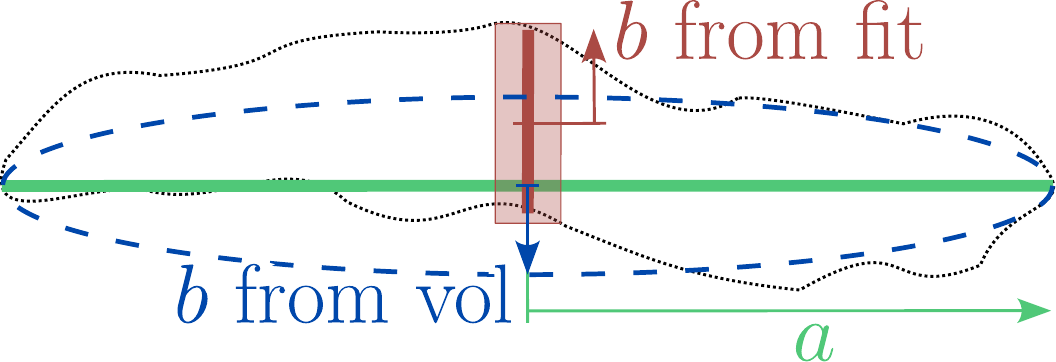}
            \label{subfig:ellipsoid_fit}%
        }
    \end{minipage}\hfill %
    \begin{minipage}[c]{0.48\linewidth}
        \centering
        \subfloat[]{%
            \includegraphics[width=\linewidth]{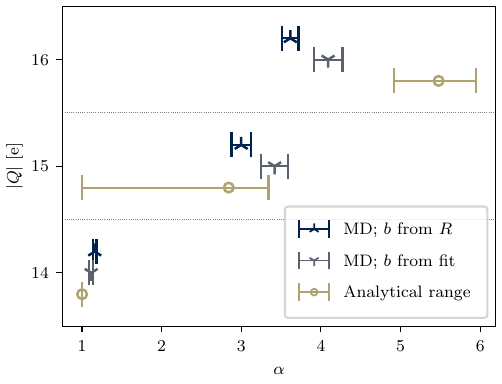}
            \label{subfig:md_comp}%
        }
    \end{minipage}
    
    \caption{(a) Equilibrium structures of charged nanoclusters with $R=2.5$ nm predicted from the \gls{md} simulations. (b) Schematic representation of the fitting process to extract the nanocluster dimensions. (c) Aspect ratio of charged nanoclusters comparing \gls{md} simulations and the analytical model.}
    \label{fig:md_equil_combined}
\end{figure*}

To validate the model, we perform \gls{md} simulations with the SPC/E water model and explicit solvated electrons at 180 K. Spherical clusters having $R=2.5$ nm and 2181 water molecules are constructed, and electrons are placed uniformly near the surface to mimic the charge distribution of a spherical conductor. Then, the charged clusters undergo successive heating and cooling stages until they spontaneously deform into ellipsoids. This successive heating-cooling approach allows us to pass the grains through quasi-equilibrium stages at sufficiently low temperatures, which allow the clusters to deform slowly into ellipsoids, rather than rapidly, leading to the formation of Taylor cones. Moreover, the clusters can deform sufficiently fast, on timescales that are accessible via \gls{md}. The \gls{md} methodology is detailed in Section S2.1. Each simulation for the 16 e and 15 e charged nanoclusters, shown in \cref{fig:md_equil_combined}, used 24 CPUs and lasted $\sim$20 days.

For the analytical model we use $\gamma_{\infty} = 79.76$ mN/m at 180 K as fitted from the SPC/E \gls{md} measurements of Ref.~\citenum{Wang2019}. A $\delta$ value close to the supercooled regime has not been reported for the SPC/E model. Recently, Ref.~\citenum{Chen2025} found $\delta \approx -0.48$ {\AA} at 300~K with measurements ranging from $-0.35$ to $-0.64$ {\AA}. Due to the absence of a specific $\delta$ value, we report analytical solutions for the whole range found in Ref.~\citenum{Chen2025}.

As the size of the cluster decreases, its charge could in principle be limited by electron field emission (EFE) because strong electrostatic repulsions cause electrons to escape by tunnelling through the potential energy barrier of the surface. In metallic spheres, EFE can in fact impose a strong current \cite{Holgate2018}. However, in water clusters, electrons get solvated and so are highly localized \cite{Kritikos2025}. In such clusters, electrons can be closer than 1 nm before electrostatic repulsions become important and lead to EFE. At larger distances, repulsions are insufficient to overcome the binding energy of electrons. Assuming that electrons are points on a spherical distribution, their minimum distance is $d_{min}=R[8\pi/(\sqrt{3}N)]^{1/2}$ \cite{Saff1997}. For our nanocluster of $R=2.5$ nm and a maximum of 17 e, $d_{min}\approx 2.3$ nm. Thus, for our nonconducting clusters, EFE is not significant.

\Cref{subfig:md_screenshots} shows the equilibrium nanoclusters obtained with \gls{md} at $\tau = 50$ ns simulation time when $|Q|=14$ e and at $\tau = 100$ ns when $|Q|>14$ e (cf. Section~S2). The elongated nanoclusters have an ellipsoidal shape, which validates our initial assumption. \Cref{subfig:md_comp} compares the aspect ratios predicted by \gls{md} and by theory. The points of the analytical model correspond to the locus of $\Delta G_{g,\min}$ points, shown in \cref{fig:energy_indiv}a, but for the conditions of the MD model. The MD averages and standard deviations are measured over $\tau \in[0,50]$ ns when $|Q|=14$ e and over $\tau \in[100,200]$ ns when $|Q|>14$ e. The aspect ratio is found with two approaches shown in \cref{subfig:ellipsoid_fit} and detailed in Section~S2.2: either by fitting both $a$ and $b$ from the nanocluster's shape (\textit{geometric} approach) or by fitting $a$ and computing $b$ from $R$ under the assumption that the nanocluster is a perfect ellipsoid (\textit{volumetric} approach). 

Both the analytical model and the MD simulations predict that the 14 e nanocluster should remain spherical (i.e., $\alpha\approx 1$) due to weak electrostatic stresses. Also, both predict a more elongated shape as the number of electrons increases. For the $|Q|=15$ e cluster, the equilibrium $\alpha$ predicted by the analytical model using $\delta=-0.48$ {\AA} agrees with both \gls{md} fitting approaches. The agreement is even better when $b$ of the \gls{md} cluster is calculated from $R$, which better aligns with the perfect ellipsoid assumption of the analytical model. As $|Q|$ increases, $\alpha$ does not increase as significantly in the \gls{md} simulations as in the analytical model. This is attributed either to the kinetic driving force being insufficient to reach the minimum energy shape predicted by the free energy analysis \cite{Kritikos2026}, or to the Tolman length not being constant and changing with the electron number and distribution, which should be investigated in future work.

The aspect ratio for the low charge case (i.e., 14 e) does not reach unity in MD simulations, as shown in \cref{subfig:md_comp}. The reason is that thermal fluctuations perturb the surface distribution, effectively leading to fluctuations of $\alpha$. For a spherical droplet, the mean square displacement of the surface due to surface oscillations is given by \cite{Henderson1978}
\begin{equation}
\langle \zeta^2 \rangle_T = \frac{1}{2 \pi} \frac{\kappa T}{\gamma} \left\{\log L + \tilde{\sigma} - \frac{11}{12} + \mathcal{O} (L^{- 1}) \right\},
\end{equation}
where $\tilde{\sigma} = 0.577$ is Euler's constant, $L = k_m R_0$, $R_0$ is the radius of the sphere at equilibrium, and $k_m$ is the maximum wavenumber cut-off for the waves, related to the number density. We can define the radii of the peaks and valleys of the oscillations as 
\begin{equation}
\begin{aligned}
R_{peaks} &= R_0  + \sqrt{\langle \zeta^2 \rangle_T} \\
R_{valleys} &= R_0  - \sqrt{\langle \zeta^2 \rangle_T}. 
\end{aligned}
\end{equation}
Thus, the aspect ratio due to thermal oscillations is $\alpha_{osc}=R_{peaks}/R_{valleys}$. At 180 K, $\alpha_{osc}\approx 1.1$, which perfectly matches the value obtained from MD averaged over the last 4 ns. This indicates that even if the equilibrium analytical model predicts a spherical nanocluster, the actual shape should be slightly elongated due to thermal fluctuations. This result also suggests that in situations where a small local $\Delta G$ maximum exists near $\alpha=1$, as seen in \cref{fig:energy_indiv}a, thermal fluctuations may be sufficient to help the nanocluster surpass this initial peak, and thus enable further elongation.

The analytical model presented here agrees well with the MD predictions, indicating that this model can capture the equilibrium structure of charged elongated nanoclusters. The model provides a comprehensive framework that can be used to predict how nanoclusters evolve in complex astrophysical and laboratory plasmas environments and in atmospheric aerosols.


\begin{acknowledgments}
PMB thanks the US National Science Foundation (NSF) for funding (grant number 2308558). WAG thanks the Liquid Sunlight Alliance (LiSA) for support, which is supported by the US Department of Energy, Office of Science, Office of Basic Energy Sciences, Fuels from Sunlight Hub under Award Number DE-SC0021266. The computations presented here were conducted in the Resnick High Performance Computing Center, a facility supported by Resnick Sustainability Institute at the California Institute of Technology.
\end{acknowledgments}


\bibliography{references}

\end{document}


\title{Equilibrium Elongation of Charged Nanoclusters -- Supporting Information}

\author{Efstratios M. Kritikos}
\email{emk@caltech.edu}
\affiliation{Department of Applied Physics and Materials Science, California Institute of Technology, Pasadena, CA 91125, United States}
\affiliation{Department of Physics, University of Oxford, Parks Road, Oxford, OX1 3PU, United Kingdom}

\author{William A. Goddard III}
\affiliation{Materials and Process Simulation Center, California Institute of Technology, Pasadena, CA 91125, United States}
\affiliation{Department of Applied Physics and Materials Science, California Institute of Technology, Pasadena, CA 91125, United States}

\author{Paul M. Bellan}
\affiliation{Department of Applied Physics and Materials Science, California Institute of Technology, Pasadena, CA 91125, United States}

\begin{abstract}
This document reports additional derivations and results in support of the discussion presented in the main paper. The structure of the Supporting Information is as follows:
\tableofcontents
\end{abstract}
\maketitle

\pagebreak

\section{Free energy of elongation}
\label{sisec:free_energy}

\subsection{Derivation of the principal radii of curvature}
\label{sisubsec:derivR1R2}

In this section, we derive the principal radii of curvature. This derivation follows closely the analysis presented in Ref. \citenum{Timoshenko1959}. As discussed in the manuscript, we consider an ellipsoid (prolate spheroid) formed from an ellipse and a revolution around the $z$ axis. The equation for the meridional ellipse is
\begin{equation}
  \frac{r^2}{b^2} + \frac{z^2}{a^2} = 1 \label{sieq:elleq}.
\end{equation}
The normal vector to the surface of the ellipsoid is parallel to the gradient
\begin{equation}
  \nabla \left( \frac{r^2}{b^2} + \frac{z^2}{a^2} \right) = \left( \frac{2
  r}{b^2}, \frac{2 z}{a^2} \right).
\end{equation}
We define the angle $\phi$ as the angle made by the normal to the ellipse and the vertical $z$-axis of revolution. Therefore, the normal vector has components proportional to
\begin{equation}
  (\sin \phi, \cos \phi).
\end{equation}
Therefore, a scalar $\xi$ exists such that the following is true
\begin{align}
    \frac{r}{b^2} &= \xi \sin \phi, \\
    \frac{z}{a^2} &= \xi \cos \phi.
  \label{sieq:xirelations}
\end{align}
Substituting Eqs.~(S4) and (S5) into \cref{sieq:elleq} gives
\begin{equation}
    \xi = (b^2 \sin^2 \phi + a^2 \cos^2 \phi)^{-1/2}
\end{equation}

Therefore, the ellipse is given with respect to the normal angle $\phi$ as
\begin{align}
    r = \frac{b^2 \sin \phi}{\sqrt{b^2 \sin^2 \phi + a^2 \cos^2 \phi}}, \\
    z = \frac{a^2 \cos \phi}{\sqrt{b^2 \sin^2 \phi + a^2 \cos^2 \phi}}.
\label{sieq:rzphi}
\end{align}  

We denote the principal radii as $R_1$ and $R_2$. These are respectively the radii of the circles sitting on the meridian plane and on the plane perpendicular to the meridian, as shown in Fig.~1 of the manuscript. A surface element on the ellipsoid has side lengths of $\dd s_1 = R_1 d\phi$ and $\dd s_2 = R_2 \sin \phi \dd \theta=r \dd \theta$.

First, we derive the principal radii $R_1$. Differentiating Eqs. (S7) and (S8) with respect to $\phi$ gives
\begin{align}
    \frac{dr}{d \phi} &= \frac{a^2 b^2 \cos \phi}{(b^2 \sin^2 \phi + a^2 \cos^2 \phi)^{3 / 2}}, \\
\frac{dz}{d \phi} &= - \frac{a^2 b^2 \sin \phi}{(b^2 \sin^2 \phi + a^2 \cos^2 \phi)^{3 / 2}}.
\end{align}
Therefore, since
\begin{align}
   \dfrac{\dd s_1}{\dd \phi} &= \sqrt{\left( \frac{d r}{d \phi} \right)^2 + \left( \frac{dz}{d \phi}\right)^2},
\end{align}
the first principal radii of curvature is
\begin{equation}
   R_1 = \frac{a^2 b^2}{(b^2 \sin^2 \phi + a^2 \cos^2 \phi)^{3 / 2}}.
\end{equation}
For $R_2$, we substitute $r$ from Eq.~(S7) into $R_2 = r/\sin \phi$ to get
\begin{equation}
    R_2 = \frac{b^2}{(b^2 \sin^2 \phi + a^2 \cos^2 \phi)^{1 / 2}}.
\end{equation}

Next, we replace the angle $\phi$ parametrization with $r$ and $z$ coordinates. From Eqs.~(S4) and (S5), we have that
\begin{equation}
    b^2 \sin^2 \phi + a^2 \cos^2 \phi = \frac{a^4 b^4}{a^4 r^2 + b^4 z^2}.
\end{equation}
This gives,
\begin{equation}
    R_1 = \frac{(a^4 r^2 + b^4 z^2)^{3 / 2}}{a^4 b^4}
    \label{sieq:R1ab}
\end{equation}
and
\begin{equation}
    R_2 = \frac{(a^4 r^2 + b^4 z^2)^{1 / 2}}{a^2},
    \label{sieq:R2ab}
\end{equation}
and the relation $R_1 = a^2 R_2^3/b^4$ holds true. We can now use the relations $a = \alpha^{2/3} R$ and $b = \alpha^{- 1 / 3} R$, to write $R_1$ and $R_2$ with respect to $\alpha$ and $R$. From \cref{sieq:elleq}, 
\begin{align}
    r &= b \sqrt{1 - \frac{z^2}{a^2}} \\
    r &= \alpha^{- 1} \sqrt{\alpha^{4 / 3} R^2 - z^2}.
\end{align}
Therefore, after substitution into \cref{sieq:R1ab,sieq:R2ab}, we get
\begin{equation}
    R_1 = \frac{1}{\alpha^{1 / 3} R^2} [\alpha^{4 / 3} R^2 + (\alpha^{- 2} - 1) z^2]^{3 / 2}
    \label{sieq:R1alphaR}
\end{equation}
and
\begin{equation}
    R_2 = \alpha^{- 1} [\alpha^{4 / 3} R^2 + (\alpha^{- 2} - 1) z^2]^{1 / 2}.
    \label{sieq:R2alphaR}
\end{equation}

\subsection{Equivalent system explanation for deformation under constant potential}
\label{sisubsec:equiv_circuit}

\begin{figure}[!htb]
    \centering
    \begin{circuitikz}
        \draw (0,0) to[battery, l=Battery] (0,3) 
            -- (4,3)                     
            to[C, l_=Capacitor] (4,0)            
            -- (0,0);                    
    
        \draw[dashed, thick] (-2, -1) rectangle (6, 4);
    \end{circuitikz}
    \caption{Equivalent circuit of a nanocluster being charged under constant potential.}
    \label{sifig:equiv_circuit}
\end{figure}
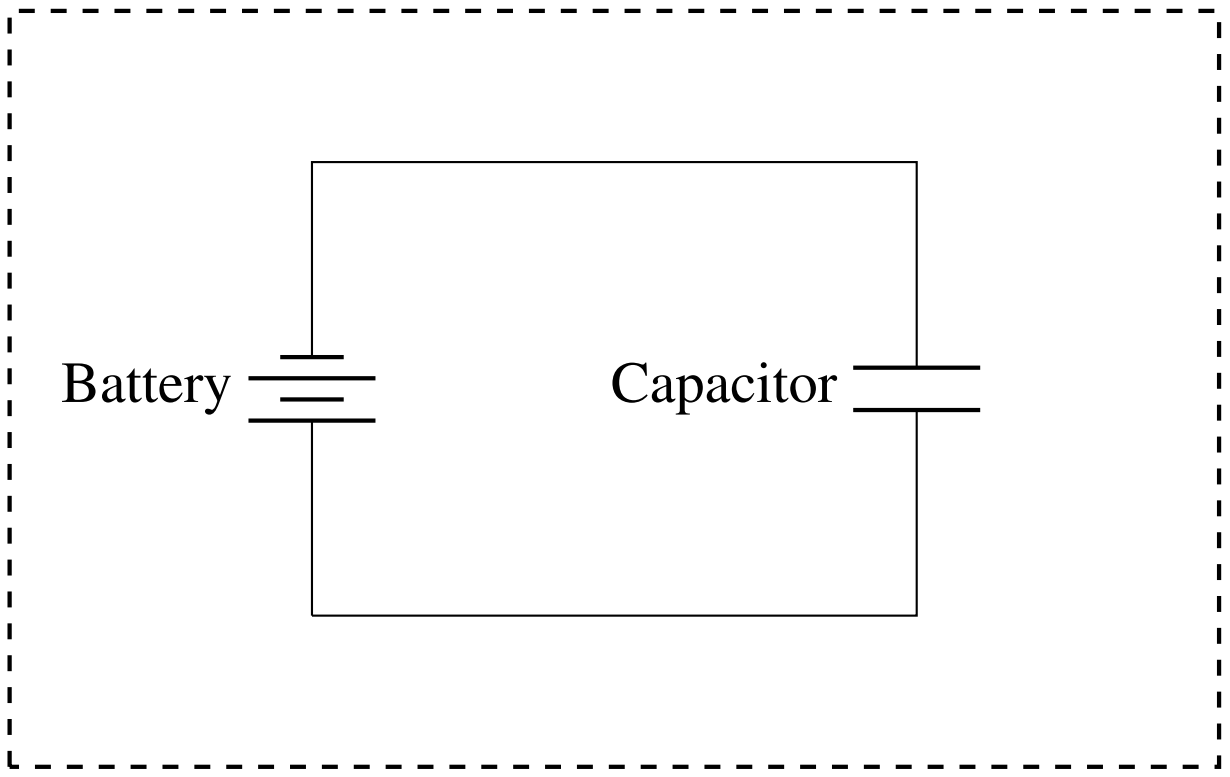

The equivalent circuit of a nanocluster being deformed under constant potential is shown in \cref{sifig:equiv_circuit}. The energy of the capacitor is 
\begin{equation}
    U=\dfrac{1}{2}C\varPhi^2.
\end{equation}
The change in the energy of the capacitor when $C\rightarrow C+\dd C$ at constant potential is 
\begin{equation}
    \dd U=\dfrac{1}{2}\varPhi^2 \dd C.
\end{equation}
The change in the capacitor charge when $C\rightarrow C+\dd C$ at constant potential is
\begin{equation}
    \dd Q = (C+\dd C) \varPhi - C \varPhi = \varPhi \dd C.
    \label{sieq:sisubsec:equiv_circuit_temp1}
\end{equation}
The external battery provides this charge. 

The associated change in the energy of the external battery $W_{\rm battery}$ is $\dd W_{\rm battery}=-\varPhi \dd Q$. Using \cref{sieq:sisubsec:equiv_circuit_temp1}, this change in energy becomes
\begin{equation}
    \dd W_{\rm battery} = -\varPhi^2 \dd C.
\end{equation}
The complete system consists of the capacitor and the external battery (cf. \cref{sifig:equiv_circuit}) and thus the energy of the complete system is 
\begin{equation}
\begin{aligned}
    \dd W_{\rm system} &= \dd U_{\rm capacitor} + \dd W_{\rm battery} \\
    & = \dfrac{1}{2}\varPhi^2 \dd C - \varPhi^2 \dd C \\
    & = -\dfrac{1}{2} \varPhi^2 \dd C.
    \label{sieq:sisubsec:equiv_circuit_temp3}
\end{aligned}
\end{equation}

Thus, the complete system will go to a lower energy state if $C$ increases. The system can increase $C$ by elongating (such as the case for the nanocluster), by having the capacitor plates come closer together (sketch in \cref{sifig:equiv_circuit} here), or by drawing in high dielectric material between the capacitor plates.

\subsection{Parametric analysis}

\begin{figure}[!htb]
    \centering
    \includegraphics[width=0.45\linewidth]{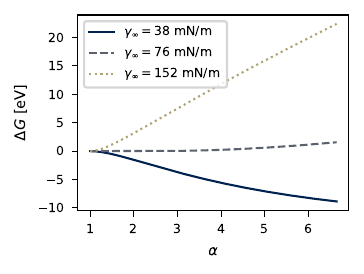}%
    \includegraphics[width=0.45\linewidth]{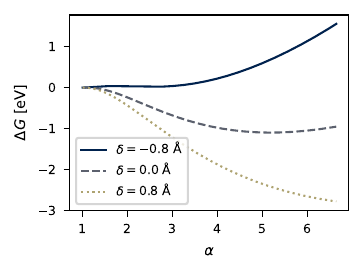}
    \includegraphics[width=0.45\linewidth]{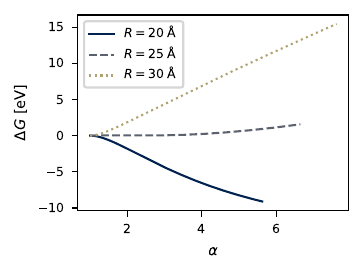}%
    \includegraphics[width=0.45\linewidth]{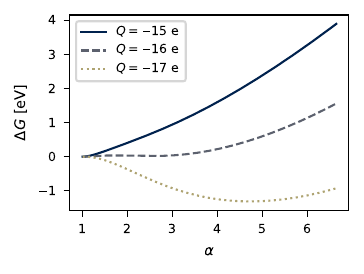}
    \caption{Parametric study of $\Delta G$ for nanocluster elongation at constant charge. The conditions are $\gamma_\infty=76$ mN/m, $\delta=-0.8$ {\AA}, $r=25$ {\AA}, $Q=-16$ e, unless otherwise specified in the legends.}
    \label{sifig:parametric_DeltaG_Q}
\end{figure}

\begin{figure}[!htb]
    \centering
    \includegraphics[width=0.45\linewidth]{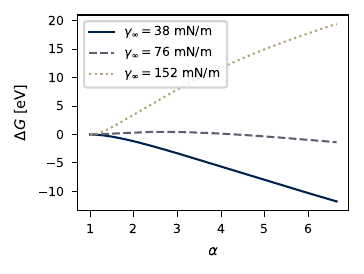}%
    \includegraphics[width=0.45\linewidth]{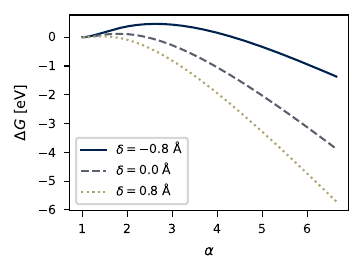}
    \includegraphics[width=0.45\linewidth]{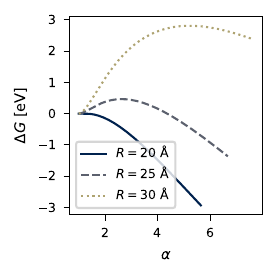}%
    \includegraphics[width=0.45\linewidth]{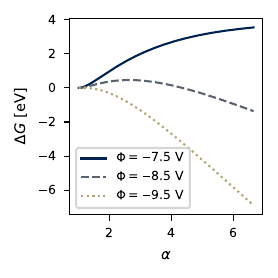}
    \caption{Parametric study of $\Delta G$ for nanocluster elongation at constant electric potential. The conditions are $\gamma_\infty=76$ mN/m, $\delta=-0.8$ {\AA}, $r=25$ {\AA}, $\varPhi=-8.5$ V, unless otherwise specified in the legends.}
    \label{sifig:parametric_DeltaG_Phi}
\end{figure}

\section{Molecular Dynamics simulations}
\label{sisec:atomistic}

\subsection{Simulation details}
\label{sisubsec:md_sim}

Simulations of water nanoclusters charged with solvated electrons are performed on the Large-scale Atomic/Molecular Massively Parallel Simulator (LAMMPS) software \cite{Thompson2022}. Water molecules are modelled with the SPC/E model \cite{Berendsen1987} and electron-water interactions with the model detailed in Ref. \citenum{Kritikos2025}. Long-range electrostatic interactions are described using the particle–particle particle-mesh (PPPM) solver. The cutoff distance of the PPPM solver is set to 10 {\AA}. Equations of motion are propagated using the Verlet algorithm with a timestep of 2 fs. The water molecules are kept rigid via the SHAKE algorithm, i.e., both the O-H bond length and H-O-H angle are kept constant throughout the simulation.

Initially, spherical nanoclusters are created with a 25 {\AA} radius and 2181 water molecules. The electrons are placed uniformly close to the surface of the cluster to mimic the charge distribution of spherical conductors. Specifically, electrons are placed on a sphere with a 23 {\AA} radius uniformly using the Fibonacci sphere sequence \cite{Marques2013}. The charged nanoclusters first undergo a geometry optimization. Next, initial velocities are assigned to the particles sampled from a Maxwell-Boltzmann distribution at 10 K. The temperature of the system is ramped up to the target temperature of 220 K using the canonical (NVT) ensemble for 40 ps. The temperature of the system is regulated using the Nos{\'e}-Hoover thermostat on the translational degrees of freedom of the particles, with a temperature damping constant of 200 fs. 

Next, simulations are performed at 220 K. The SPC/E model predicts a lower melting point of 215 K compared to experimental measurements \cite{Vega2005}; therefore, at 220 K, we are above the melting point, and thus accelerated kinetics are expected. Simulating at a lower temperature increases the computational cost significantly, making the simulations prohibitively time-consuming. Due to the accelerated kinetics at these elevated temperatures, the nanoclusters either form protrusions and the electrons swiftly escape or start elongating and break apart. To mimic the low-temperature conditions of astrophysical nanoclusters and successfully slow down the kinetics of the cluster as it elongates, a freezing step is included. Specifically, we monitor the radius of gyration of the cluster $r_g^2=\sum_i m_i (\bm{r}_i-\bm{r}_{com})^2/M$, where $m_i$ is the mass of atom $i$, $M$ is the total mass of the cluster, $\bm{r}$ is the position vector, and $\bm{r}_{com}$ is the center-of-mass position of the cluster. 

Then, the following condition is applied: if $r_g$ increases by more than 15\% or the rate of change of $r_g$ is more than 250 {\AA}/ns, the cluster is frozen to 10 K, energy equilibrated for 20 ps, and then gradually reheated to 220 K over a period of 100 ps. Going through these successive quasi-equilibrium states guarantees that the structure slowly evolves, mimicking a much lower temperature condition. Elongated structures were created with this process, which had an average temperature of $\sim 180$ K. Structures that formed protrusions were discarded from the analysis.

\begin{figure}[!t]
    \centering
    \includegraphics[width=0.5\linewidth]{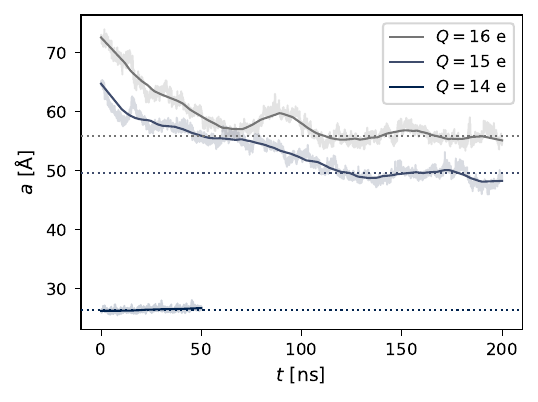}%
    \includegraphics[width=0.5\linewidth]{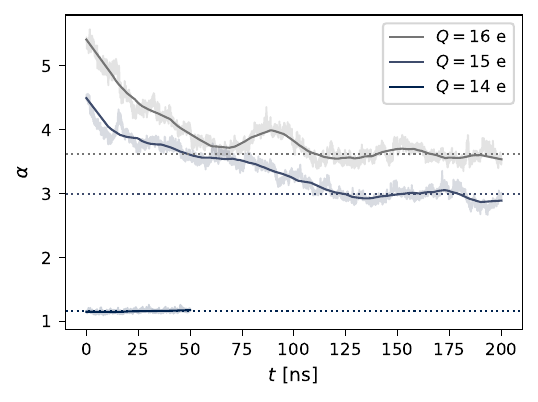}
    \caption{Dimensions of nanoclusters during the final equilibration stage at 180 K.}
    \label{sifig:MD_time_equil}
\end{figure}

The extracted structures are optimized again, assigned velocities at 10 K, heated from 10 K to 150 K over 40 ps, and then energy equilibrated at 150 K for at least 15 ns. Afterwards, they are heated to 180 K over 40 ps (which temperature is equal to the average temperature of the initial simulations), and then are simulated at constant temperature for 200 ns. The time evolution of the aspect ratio $\alpha$ and radius $a$ of the nanoclusters is shown in \cref{sifig:MD_time_equil}. 

All simulations were run on 24 CPUs, reaching an average of $\sim$13 ns/day. The nanoclusters with 15 e and 16 e were simulated for a total of $\sim 255$ ns, i.e., $\sim$40 ns in the freezing and reheating stage at 220~K, 15 ns in the 150 K stage, and 200~ns in the 180~K stage. Each one of those simulations lasted $\sim$20 days.

\subsection{Extracting cluster dimensions from molecular dynamics simulations}
\label{subsec:clusterdims_MDsims}

The dimensions of the nanoclusters are computed from the surface area. The cluster surface is constructed following the alpha-shape method \cite{Stukowski2014}, in which a spherical probe is used to trace the surface atoms, followed by an iterative smoothing scheme. The OVITO software \cite{Stukowski2010} is used for this computation, using a probe sphere with a 10 {\AA} radius and a smoothing level of 8. 

After the surface is constructed, the pair of points on the surface with the largest distance is identified. The major radius $a$ is set equal to half the distance between these two surface points. A schematic representation is shown in Fig.~4b of the manuscript. Then, two approaches are used to compute the minor radius $b$: (i) using the assumption that the cluster is a perfect ellipsoid (\textit{volumetric} approach), or (ii) by directly measuring the radius $b$ at the midpoint (\textit{geometric} approach). In case (i), the volume within the constructed surface, $V$, is computed. $R$ is found from $R^3=3 V/(4\pi)$ and then the dimensions $b^2=R^3/a$ and $\alpha=a/b$ are computed. In case (ii), first the midpoint along the line between the two furthest points of the surface is found, and the plane passing through the midpoint with a normal vector parallel to $a$ is determined. Then, all surface points within a distance of 2 {\AA} from the midpoint plane are selected and projected onto the plane. Finally, the centroid is computed, and $b$ is determined from the average distance between the projected points and the centroid. Case (i) is directly related to the theory developed here, whereas case (ii) is a direct measurement of the dimensions of the nanoclusters. When the computed dimensions of cases (i) and (ii) match, it means that the cluster is very close to an ellipsoidal shape.

\bibliography{references}